\documentclass{emulateapj}

\usepackage{graphicx}
\usepackage[]{natbib}
\usepackage{placeins}
\usepackage{lineno}
\usepackage{ulem}
\usepackage{rotating}
\usepackage{adjustbox}
\usepackage{color}

\usepackage[caption=false]{subfig}    

\pdfoutput=1

\newcommand{\ltsima} {$\; \buildrel < \over \sim \;$}

\newcommand{\gtsima} {$\; \buildrel > \over \sim \;$}
\newcommand{\lta} {\lower.5ex\hbox{\ltsima}}
\newcommand{\gta} {\lower.5ex\hbox{\gtsima}}

\shorttitle{A Unified Model for GRB Prompt Emission from Optical to $\gamma$-rays; Exploring GRBs as Standard Candles}
\shortauthors{S.Guiriec}

\begin{document}

\title{A Unified Model for GRB Prompt Emission from Optical to $\gamma$-rays;  Exploring GRBs as Standard Candles}

\author{S. Guiriec\altaffilmark{1,2,3,4}, C. Kouveliotou\altaffilmark{1}, D. H. Hartmann\altaffilmark{5}, J. Granot\altaffilmark{6}, K. Asano\altaffilmark{7}, P. M\'esz\'aros\altaffilmark{8}, R. Gill\altaffilmark{6}, N. Gehrels\altaffilmark{2} \& J. McEnery\altaffilmark{2}}

\altaffiltext{1}{Department of Physics, The George Washington University, 725 21st Street NW, Washington, DC 20052, USA}
\altaffiltext{2}{NASA Goddard Space Flight Center, Greenbelt, MD 20771, USA}
\altaffiltext{3}{Department of Physics and Department of Astronomy, University of Maryland, College Park, MD 20742, USA}
\altaffiltext{4}{Center for Research and Exploration in Space Science and Technology (CRESST)}
\altaffiltext{5}{Department of Physics and Astronomy, Clemson University, Kinard Lab of Physics}
\altaffiltext{6}{Department of Natural Sciences, The Open University of Israel, 1 University Road, P.O. Box 808, RaÕanana 4353701, Israel}
\altaffiltext{7}{Institute for Cosmic Ray Research, The University of Tokyo, 5-1-5 Kashiwanoha, Kashiwa, Chiba 277-8582, Japan}
\altaffiltext{8}{Department of Astronomy \& Astrophysics and Department of Physics, Center for Particle and Gravitational Astrophysics, Pennsylvania State University, University Park, PA 16802, USA}

\email{sylvain.guiriec@nasa.gov}

\begin{abstract}
The origin of prompt emission from gamma ray bursts remains to be an open question. Correlated prompt optical and $\gamma$-ray emission observed in a handful of GRBs strongly suggests a common emission region, but failure to adequately fit the broadband GRB spectrum prompted the hypothesis of different emission mechanisms for the low- and high-energy radiations. We demonstrate that our multi-component model for GRB $\gamma$-ray prompt emission provides an excellent fit to GRB~110205A from optical to $\gamma$-ray energies. Our results show that the optical and highest $\gamma$-ray emissions have the same spatial and spectral origin, which is different from the bulk of the X- and softest $\gamma$-ray radiation. Finally, our accurate redshift estimate for GRB~110205A demonstrates promise for using GRBs as cosmological standard candles.

\end{abstract}


\section{Introduction}

During their prompt emission phase, gamma-ray bursts (GRBs) produce the most luminous electromagnetic flashes in Nature that are seen across most of the visible universe. The $\gamma$-ray emission arises from violent energy-dissipation mechanisms within a highly relativistic jet powered by a newly-born stellar-mass black hole or rapidly rotating highly-magnetized neutron star~(see for e.g., \citet{Piran:2005,Kumar:2015} for reviews; \citet{Shemi:1990,Rees:1992,Rees:1994,Meszaros:1993}). Decades after the first GRB detection, the nature of this $\gamma$-ray emission remains an outstanding question. It has been traditionally associated to synchrotron emission from charged particles accelerated within the jet~\citep[e.g.,][]{Rees:1994,Kobayashi:1997,Daigne:1998,Daigne:2002,Daigne:2011,Zhang:2011} and/or to reprocessed thermal radiation field emanating from the jet photosphere~\citep[e.g.,][]{Meszaros:2000,Eichler:2000,Peer:2006,Beloborodov:2010,Vurm:2011,Lazzati:2013,Lazzati:2016}.

In a few dozen cases, optical emission has been detected during the $\gamma$-ray prompt phase~\citep[e.g.,][]{Akerlof:1999,Fox:2003,Blake:2005,Vestrand:2005,Vestrand:2006,Racusin:2008,Cucchiara:2011,Gendre:2012,Zheng:2012}. Such a multi-wavelength coverage is instrumental for understanding the underlying prompt-emission mechanisms. The optical and $\gamma$-ray correlated variability observed in a few GRBs pointed toward a common spatial origin within the jet~\citep[e.g.,][]{Vestrand:2005,Vestrand:2006,Racusin:2008,Cucchiara:2011,Gendre:2012,Zheng:2012}. In addition, the constant $\gamma$-ray to optical flux ratio measured in several GRBs~\citep{Vestrand:2005,Vestrand:2006} suggested that the two energy regimes are not independent. However, no $\gamma$-ray prompt emission spectral models, including the traditional Band function~\citep{Band:1993} or its variant with a high-energy cutoff has succeeded so far in fitting the broadband spectrum~\citep{Racusin:2008,Cucchiara:2011,Gendre:2012,Zheng:2012} suggesting that the emissions in the two energy ranges originate from different processes.

Recent analysis of $\gamma$-ray prompt emission of GRBs detected with the Fermi Gamma-ray Space Telescope (Fermi) as well as stored in the archival data of the Burst And Transient Source Experiment (BATSE) on board the Compton Gamma-Ray Observatory (CGRO) revealed the contribution of several separate components to the total $\gamma$-ray spectrum~\citep[e.g.,][]{Guiriec:2010,Guiriec:2011,Guiriec:2013,Guiriec:2015a,Guiriec:2015b,Guiriec:2016}. Here, we apply this new spectro-temporal multi-component model~\citep{Guiriec:2015a,Guiriec:2016} to the simultaneous data of four instruments covering the early phase of GRB~110205A from optical to MeV energies. We show that the optical and highest $\gamma$-ray emissions belong to the same spectral component that extends across six energy decades and that most of the X- and soft $\gamma$-ray emission arises from the modelÕs two other components with a different spatial origin. This unified model reveals the various components involved in the prompt emission, thereby opening a brand-new window towards (i) understanding of the nature and origin of GRB prompt emission and the underlying emission and dissipation mechanisms powering GRBs, and (ii) the use of GRBs as cosmological standard candles.

\begin{figure*}[ht!]
\begin{center}
\includegraphics[totalheight=0.50\textheight, clip, viewport=0 371 610 785]{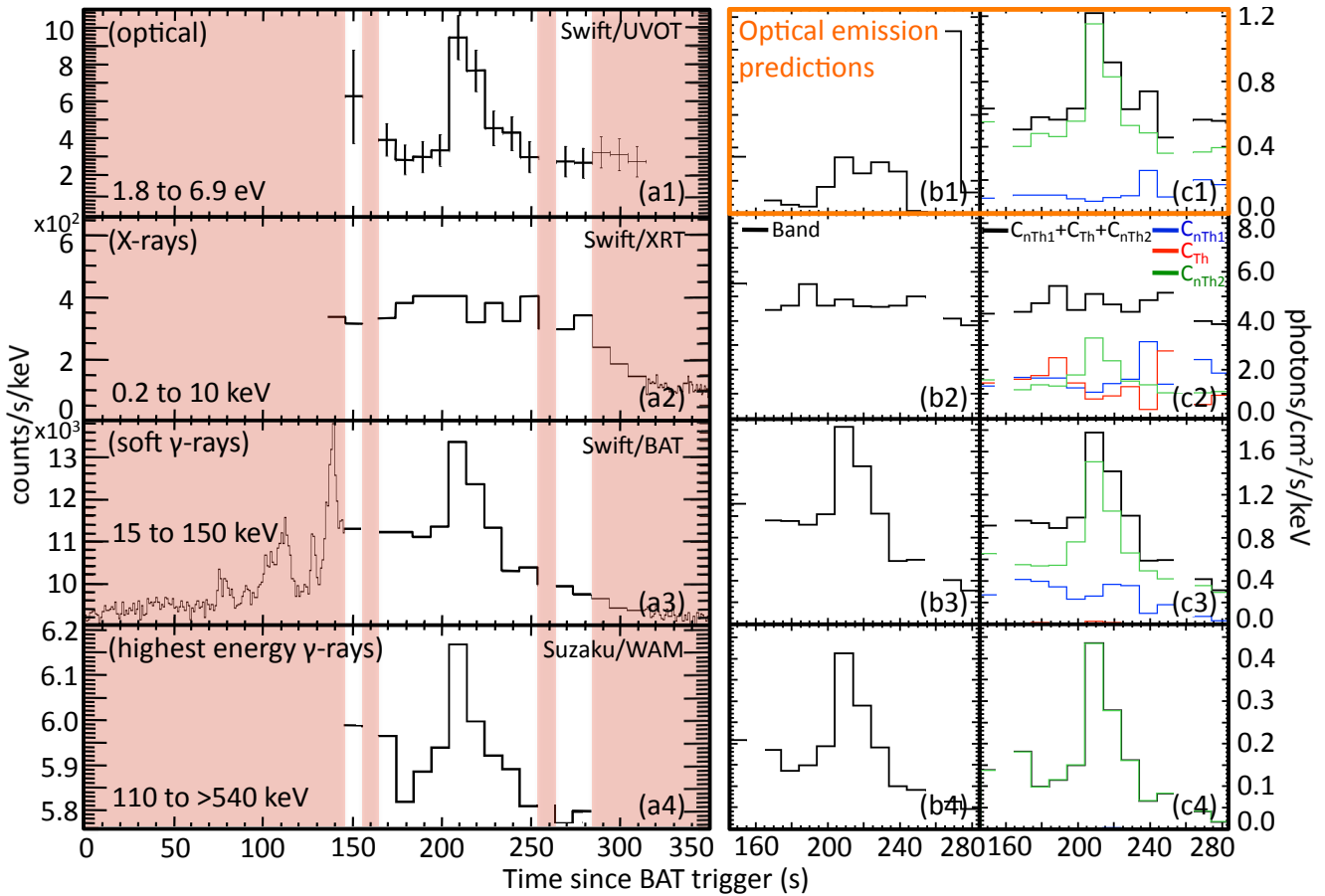}
\caption{\label{fig01}a- GRB 110205A light curves as recorded with the four instruments. The X- and $gamma$-ray light curves were modified to match with the optical time-intervals where the data of the four instruments overlap. The red shaded regions were excluded from the analysis; a \& b- Light curves resulting from the Band function (b) and C$_{nTh1}$+C$_{Th}$+C$_{nTh2}$ (c) fits to the X- and $\gamma$-ray data only in the same energy bands as the four instruments, and extrapolation into the optical regime (b1 and c1) for comparison with the observations (a1).}
\end{center}
\end{figure*}

\vspace{1.0cm}
\section{Observation}

GRB~110205A triggered the Burst Alert Telescope (BAT) on board NASA's Swift observatory on 2011, February 5, at  T$_0$$=$02:02:41 UT~\citep{Beardmore:2011}. The spacecraft repointed its X-Ray and Ultraviolet/Optical Telescopes (XRT, UVOT) to the burst direction and was able to detect X-ray and optical counterparts, starting at  T$_0$$+$146 s, simultaneously with the $\gamma$-rays, owing to the particularly long $\gamma$-ray duration of $>$300 s (Figure~\ref{fig01}a). The burst was also detected with the Wide-band All-sky Monitor (WAM) onboard the Suzaku observatory, complementing the BAT $\gamma$-ray coverage above 150 keV up to 5 MeV~\citep{Sugita:2011}. Subsequent observations of the GRB optical spectrum with the FAST spectrograph on the Whipple Observatory 1.5 m telescope determined its redshift to be z$=$2.22~\citep{Cenko:2011}, corresponding to a look-back time of 10.7 billion years (for a standard cosmology, [$\Omega_\Lambda$, $\Omega_M$, $h$]$=$[0.73, 0.27, 0.71]), only 3 billion years after the big bang.

An exceptional characteristic of this event is the correlated optical and $\gamma$-ray variability during its prompt emission, illustrated by a very intense and narrow pulse at $\sim$T$_0$$+$210 s, present in the optical and $\gamma$-ray wavelengths (Figure~\ref{fig01}a); conversely, the X-ray emission remains roughly constant over the entire burst duration. The presence of such features has thus far not been successfully modeled over a very broad spectral range~\citep{Cucchiara:2011,Gendre:2012,Zheng:2012}.

\section{Analysis}

\begin{table*}
\caption{\label{tab01}Values of the spectral indices and $\chi^2$ resulting from the fits to XRT, XRT+BAT+WAM and UVOT+XRT+BAT+WAM data in all time intervals. XRT data are fitted to a single PL or to the combination of a PL with either a black body or a CPL. XRT+BAT+WAM and UVOT+XRT+BAT+WAM data are fitted to a Band function, a Band function with a high-energy exponential cutoff and C$_{nTh1}$$+$C$_{Th}$+C$_{nTh2}$. When possible, the spectral index of C$_{nTh2}$, $\alpha_{nTh2}$, is left free to vary in the fits to UVOT+XRT+BAT+WAM data.
}
\begin{center}
\scalebox{0.65}{
\begin{tabular}{c|cccc|cccccc|ccccccc}

Time & \multicolumn{4}{c|}{XRT} & \multicolumn{6}{c|}{XRT+BAT+WAM} & \multicolumn{7}{c}{UVOT+XRT+BAT+WAM} \\
since T$_0$ & \multicolumn{4}{c|}{} & \multicolumn{6}{c|}{} & \multicolumn{7}{c}{} \\
\hline
 & PL & PL+BB & \multicolumn{2}{c|}{PL+CPL} & \multicolumn{2}{c}{Band} & \multicolumn{2}{c}{Band+cutoff} & \multicolumn{2}{c|}{C$_{nTh1}$+C$_{Th}$+C$_{nTh2}$} & \multicolumn{2}{c}{Band} & \multicolumn{2}{c}{Band+cutoff} & \multicolumn{3}{c}{C$_{nTh1}$+C$_{Th}$+C$_{nTh2}$} \\
\hline
 & $\chi^2$/dof & $\chi^2$/dof & CPL index & $\chi^2$/dof & $\alpha_{Band}$ & $\chi^2$/dof & $\alpha_{Band}$ & $\chi^2$/dof & $\alpha_{nTh2}$ & $\chi^2$/dof & $\alpha_{Band}$ & $\chi^2$/dof & $\alpha_{Band}$ & $\chi^2$/dof & $\alpha_{nTh2}$ & $\chi^2$/dof & $\alpha_{nTh2}$ \\
\hline
 146--155 s & 48/50 & 35/48 & +0.69$^{+1.13}_{-0.19}$ & 35/47 & -0.64$\pm$0.69 & 144/141 & -0.44$\pm$0.16 & 127/140 & -1 & 127/139 & -0.68$\pm$0.08 & 148/142 & -0.49$\pm$0.15 & 133/141 & -1 & 129/140 & -1 \\

 164--174 s & 79/57 & 56/55 & +0.58$^{+0.91}_{-0.23}$ & 55/54 & -0.45$\pm$0.09 & 173/148 & -0.27$\pm$0.14 & 167/147 & -1 & 166/146 & -0.62$\pm$0.07 & 187/149 & -0.62$\pm$0.07 & 184/148 & -1 & 166/147 & -1.09$\pm$0.03 \\
 
 174--184 s & 83/63 & 62/61 & +0.51$^{+0.87}_{-0.15}$ & 60/60 & -0.39$\pm$0.08 & 156/154 & -0.17$\pm$0.12 & 141/153 & -1 & 141/152 & -0.44$\pm$0.07 & 166/155 & -0.44$\pm$0.07 & 152/154 & -1 & 140/153 & -0.96$\pm$0.04 \\
 
 184--194 s & 81/62 & 61/60 & +0.52$^{+0.93}_{-0.15}$ & 61/59 & -0.28$\pm$0.11 & 151/153 & -0.22$\pm$0.14 & 146/152 & -1 & 144/151 & -0.34$\pm$0.10 & 162/154 & -0.34$\pm$0.10 & 158/153 & -1 & 144/152 & -0.99$\pm$0.10 \\
 
 194--204 s & 58/58 & 46/56 & +0.54$^{+0.97}_{-0.03}$ & 45/55 & -0.58$\pm$0.08 & 163/149 & -0.41$\pm$0.13 & 141/148 & -1 & 141/147 & -0.70$\pm$0.05 & 169/150 & -0.70$\pm$0.05 & 152/149 & -1 & 141/148 & -0.97$\pm$0.04 \\
 
 204--214 s & 53/59 & 41/57 & +0.56$^{+0.95}_{-0.07}$ & 41/56 & -0.72$\pm$0.06 & 193/150 & -0.30$\pm$0.19 & 108/149 & -1 & 107/146 & -1.02$\pm$0.01 & 176/151 & -1.02$\pm$0.01 & 141/150 & -1 & 123/149 & -1 \\
 
 214--224 s & 72/57 & 61/55 & +1.14$^{+1.23}_{-0.56}$ & 61/54 & -0.67$\pm$0.06 & 189/148 & -0.52$\pm$0.11 & 146/147 & -1 & 144/146 & -0.95$\pm$0.02 & 201/149 & -0.95$\pm$0.02 & 174/148 & -1 & 148/147 & -1.06$\pm$0.03 \\
 
 224--234 s & 65/55 & 43/53 & +1.10$^{+1.14}_{-0.70}$ & 43/52 & -0.70$\pm$0.07 & 160/146 & -0.37$\pm$0.17 & 149/145 & -1 & 143/144 & -0.83$\pm$0.04 & 166/147 & -0.83$\pm$0.04 & 161/146 & -1 & 143/145 & -1.07$\pm$0.03 \\
 
 234--244 s & 43/50 & 38/48 & +0.80$^{+1.22}_{-0.11}$ & 38/47 & -0.63$\pm$0.13 & 121/141 & -0.66$\pm$0.12 & 121/140 & -1 & 122/139 & -0.81$\pm$0.04 & 124/142 & -0.81$\pm$0.04 & 125/141 & -1 & 122/140 & -0.98$\pm$0.05 \\
 
 244--254 s &   7/11 &     2/9 & +0.51$^{+0.89}_{-0.14}$ &     2/8 & -0.08$\pm$0.34 &   96/102 & -0.08$\pm$0.38 &   95/101 & -1 &    88/100 & -0.69$\pm$0.07 & 105/103 & -0.69$\pm$0.07 & 104/102 & -1 &   89/101 & -1 \\  
 
 264--274 s & 63/47 & 53/45 & +0.74$^{+0.81}_{-0.09}$ & 54/44 & -0.92$\pm$0.11 & 168/138 & -0.85$\pm$0.15 & 166/137 & -1 & 160/136 & +0.12$\pm$0.07 & 201/139 & +0.12$\pm$0.07 & 166/138 & -1 & 160/137 & -1 \\
 
 274--284 s & 91/48 & 63/46 & +0.31$^{+0.48}_{-0.13}$ & 57/45 & -0.49$\pm$0.17 & 152/139 & -0.46$\pm$0.19 & 150/138 & -1 & 148/137 & -0.71$\pm$0.07 & 155/140 & -0.71$\pm$0.07 & 153/139 & -1 & 148/138 & -1 \\

\end{tabular}
}
\end{center}
\end{table*}

\begin{figure}[ht!]
\begin{center}
\includegraphics[totalheight=0.46\textheight, clip, viewport=0 379 323 792]{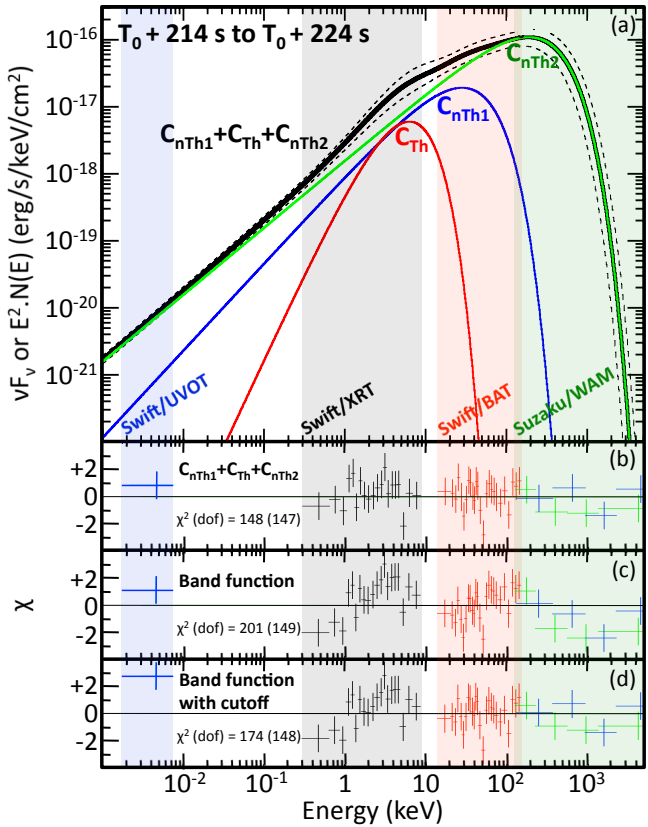}
\caption{\label{fig02}a- C$_{nTh1}$+C$_{Th}$+C$_{nTh2}$ fit to the four instrument data from T$_0$+ 214 s to T$_0$+ 224 s (solid black line) with the 1-$\sigma$ confidence region (dashed lines); b--d- Residuals of the fits using C$_{nTh1}$+C$_{Th}$+C$_{nTh2}$ (b), a Band function (c) and a Band function with a high-energy exponential cutoff (d). The energy channels have been combined for display purpose only. The resulting $\chi^2$ values of the fits are also indicated together with the number of degrees of freedom (dof).}
\end{center}
\end{figure}

We performed a time-resolved spectral analysis (from T$_0$$+$146 s to T$_0$$+$284 s) on the simultaneous data of all four instruments (Swift/UVOT, XRT, BAT and Suzaku/WAM) using $\sim$10 s time intervals to match the resolution of the optical data bins (Figure~\ref{fig01}a). When fitting X-ray data, we convolved the spectral models with attenuation from the Milky Way and the host galaxy using hydrogen column densities, N$_H$, fixed at 1.6$\times$10$^{20}$ cm$^2$ and 4.0$\times$10$^{21}$ cm$^2$, respectively~\citep{Zheng:2012,Kalberla:2005}. For the spectral fits including optical data, we convolved the spectral models with dust extinction from the Milky Way with E(B-V)=0.01~\citep{Schlegel:1998} and from the host galaxy assuming a Small Magellanic Cloud (SMC) density with E(B-V)=0.08~\citep{Zheng:2012}. 

We compared the results obtained with our new three-component model~\citep{Guiriec:2015a,Guiriec:2016} to those obtained with the traditional single-component models for $\gamma$-ray prompt emission: the Band function and the Band function with an exponential high-energy cutoff (Figures~\ref{fig01} and \ref{fig02}); the Band function is a smoothly broken power law defined with four free parameters, the indices of the low- and high-energy power laws, $\alpha_{Band}$ and $\beta_{Band}$, the spectral break energy, E$_{peak}^{Band}$, and a normalization parameter~\citep{Band:1993}. The three-component model (hereafter, C$_{nTh1}$$+$C$_{Th}$+C$_{nTh2}$) is composed of a thermal-like (C$_{Th}$) and two non-thermal (C$_{nTh1}$ and C$_{nTh2}$) components~\citep[][and Figure~\ref{fig02}a]{Guiriec:2015a,Guiriec:2016}. All three are adequately described by power laws with exponential cutoffs, with photon spectral index $\alpha$$=$$d~log(d N_{ph}/d E_{ph})/d~log(E_{ph})$ fixed to $\alpha_{nTh1}$$=$-0.7, $\alpha_{Th}$$=$+0.6 and $\alpha_{nTh2}$$=$-1.0, respectively, and with normalization parameters and spectral break energies, E$^{nTh1}_{peak}$, E$^{Th}_{peak}$ and E$^{nTh2}_{peak}$ left free to vary. Despite the larger number of component, C$_{nTh1}$$+$C$_{Th}$+C$_{nTh2}$ has only two more free parameters than the Band function (six compared to four), and one more free parameter than the Band function with a high-energy exponential cutoff; the probability that the statistical improvement obtained with C$_{nTh1}$$+$C$_{Th}$+C$_{nTh2}$ compared to the other two models is due to signal and/or background fluctuations is $<$10$^{-6}$~\citep[see simulation procedure in][]{Guiriec:2015a}. In all time intervals, the three-component model leads to the best fits (Table~\ref{tab01}); it is particularly evident in Figure~\ref{fig02} where C$_{nTh1}$$+$C$_{Th}$+C$_{nTh2}$ is the only good fit. When the parameter $\alpha_{nTh2}$ is left free to vary in the various time intervals, its values cluster around -1 as proposed in~\citet{Guiriec:2015a} (Table~\ref{tab01}).

\begin{figure*}[ht!]
\begin{center}
\includegraphics[totalheight=0.30\textheight, clip, viewport=0 5 600 240]{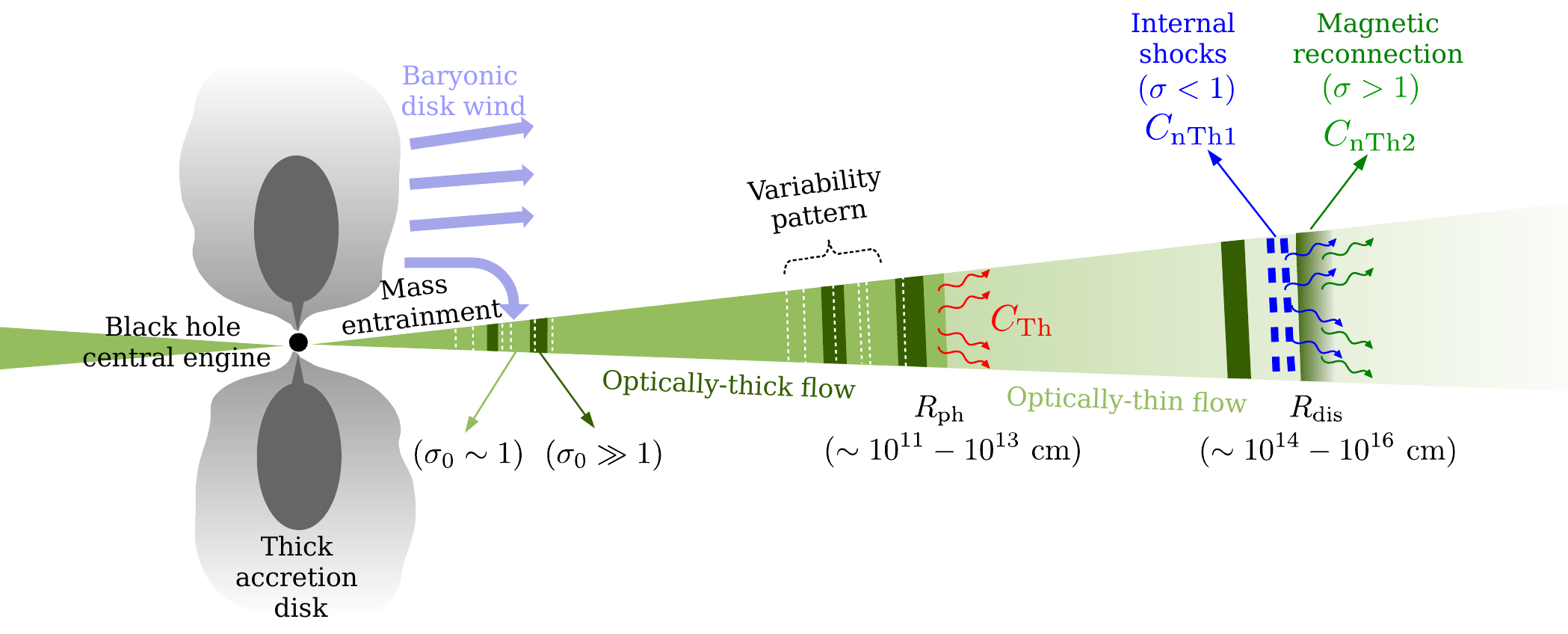}
\caption{\label{fig03}An illustration of our proposed origin of the three components:  C$_{Th}$---quasi-thermal photospheric emission, and C$_{nTh1}$ \& C$_{nTh2}$---non-thermal emission. The light and dark green bands represent low- and high-$\sigma$ (or magnetization) regions, respectively. The central engine's intermittent activity and mass entrainment from the baryonic accretion-disk wind together embed variability in the flow in the form of low- and high-$\sigma$ regions and also slower and faster moving shells, shown as white dotted lines. The flow remains optically thick until the photospheric radius $R_{ph}$ where quasi-thermal radiation (C$_{Th}$) escapes. At the dissipation radius $R_{dis}$, collisions between low-$\sigma$ outflowing shells produce internal shocks (shown as the thick dashed blue lines) that give rise to C$_{nTh1}$, while magnetic reconnection in high-  regions gives rise to C$_{nTh2}$.}
\end{center}
\end{figure*}

In the context of C$_{nTh1}$$+$C$_{Th}$+C$_{nTh2}$, both the optical and highest $\gamma$-ray emissions are related to C$_{nTh2}$, which extends across the whole spectrum over six energy decades (Figures~\ref{fig01}c \& \ref{fig02}). We tested the robustness of C$_{nTh1}$$+$C$_{Th}$+C$_{nTh2}$ as a broadband prompt emission model by investigating its ability to predict the optical emission from the analysis of high-energy data only. To do so, we fitted only the X- and $\gamma$-ray data and then extrapolated the fits to the optical regime; for comparison purposes, we performed the same exercise with the other models (Table~\ref{tab01} and Figures~\ref{fig01}b \& c). The three-component model reproduces in great details both the variability and the absolute flux of the observed light curves, conversely to the other models, which systematically under-predict the optical fluxes and do not produce the correlated optical and $\gamma$-ray variability. 
This is well illustrated by the strong similarity of the spectral parameter values and the goodness of the fits resulting from the C$_{nTh1}$$+$C$_{Th}$+C$_{nTh2}$ fits to the data from either optical up to $\gamma$-ray or to X- and $\gamma$-ray data only; in contrast, the spectral parameter values are significantly different when comparing the other model fits to either the broadband spectrum or to the high-energy data only (Table~\ref{tab01}).
The spectral fits to the X-ray data only reveal the existence of a component in addition to the PL typically used to fit this energy regime. While the PL accounts for the contributions of both C$_{nTh1}$ and C$_{nTh2}$, this additional component has the expected thermal shape of C$_{Th}$ and it is adequately described with a black body or a CPL with a positive spectral index (Table~\ref{tab01})\footnote{A similar thermal-like component in {\it Swift}/XRT data has also been recently reported in~\citet{Nappo:2016}.}.

Figures~\ref{fig01}c \& \ref{fig02}a shows the contribution of each component to the GRB light curves in the energy range of each instrument. There is clear evidence that C$_{nTh2}$ accounts for both the optical and the highest energy $\gamma$-ray emission, while the combined variability of all three components results in an overall flat X-ray light curve. The identification of these separate emission contributions is the crucial step for disentangling the simultaneous emission processes taking place within GRB relativistic jets and for finally identifying the origin of the prompt optical emission. The common spectral origin of the lowest and highest energy parts of the GRB~110205A spectrum indicates that they are related to a single emission process in direct contrast with previous conclusions~\citep{Cucchiara:2011,Gendre:2012,Zheng:2012}.  Moreover, the lack of correlated variability between C$_{nTh2}$ and the two other components (Figure~\ref{fig01}c) indicates that these originate from different regions in the outflow.

\section{Interpretation}

Below, we proceed to discuss the nature of these three components and their relation to the GRB prompt emission physics. Figure~\ref{fig03} illustrates our interpretation of the C$_{nTh1}$$+$C$_{Th}$+C$_{nTh2}$ model. C$_{Th}$'s quasi-thermal spectral shape strongly suggests a photospheric origin. Its slightly softer low-energy photon index ($\alpha_{Th}$$\sim$+0.6) compared to a pure black body spectrum ($\alpha_{Th}$$\sim$+1) may naturally arise from the superposition of different temperatures from different angles to the line of sight corresponding to different Doppler factors~\citep[e.g.,][who obtains $\alpha_{Th}$$\sim$+0.5]{Beloborodov:2010}.

The typical partial temporal correlation and slight delay of C$_{nTh1}$ with respect to C$_{Th}$~\citep{Guiriec:2011,Guiriec:2013,Guiriec:2015a,Guiriec:2015b,Guiriec:2016} naturally occurs if C$_{nTh1}$ originates in internal shocks~\citep[e.g.,][]{Rees:1994,Kobayashi:1997,Daigne:1998}, within the same outflowing plasma, that occur at a larger radius (distance from the central source) $R_{IS}$ compared to the photospheric radius $R_{ph}$. In this picture their temporal correlation arises from the variability pattern embedded into the flow by the central engineÕs intermittent activity, while C$_{nTh1}$'s slight delay is due to the slower than light speed of the bulk flow causing photons emitted at $R_{ph}$ to reach $R_{IS}$ before the flow. C$_{nTh1}$'s spectrum is suggestive of synchrotron emission from a nearly mono-energetic or thermal electron energy distribution, for which a photon index of $\alpha_{nTh1}$$=$-2/3$\sim$-0.7 is expected. Efficient internal shocks within the outflow require a relatively low magnetization parameter (magnetic-to-particle energy ratio), $\sigma$$<$1. However, unless $\sigma$$<$10$^{-3}$, which is highly unlikely because of the large-scale ordered magnetic field advected from near the central source, this magnetic field is strong enough to suppress diffusive shock acceleration~\citep{Sironi:2009,Sironi:2011}, and results in a quasi-thermal nearly mono-energetic electron energy distribution. If the electron cooling is slow (compared to the dynamical time) this leads to a C$_{nTh1}$-like spectrum. The latter may also be produced for fast-cooling electrons if they are continuously heated, e.g. via turbulence~\citep{Asano:2015,Lyubarsky:2009,Komissarov:2009}, leading to a balance between electron heating and cooling.

\begin{figure*}[ht!]
\begin{center}
\includegraphics[totalheight=0.30\textheight, clip, viewport=0 545 615 790]{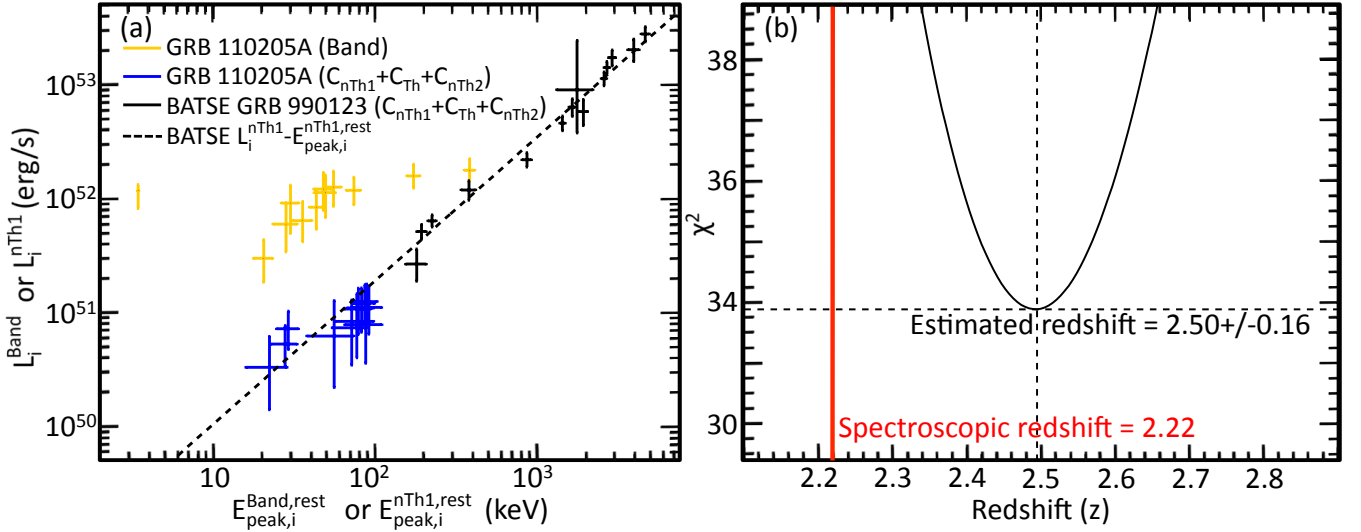}
\caption{\label{fig04}a- Time-resolved luminosities, L$_i^{nTh1}$, as a function of the time-resolved rest-frame E$_{peak,i}^{rest}$ when fitting a Band function to GRB 110205A (yellow) or C$_{nTh1}$+C$_{Th}$+C$_{nTh2}$ to GRB~110205A (blue) and BATSE GRB~990123 (black). The dashed black line corresponds to the L$_i^{nTh1}$--E$_{peak,i}^{nTh1,rest}$ relation derived from a sample of BATSE GRBs with known redshifts; b- ${\chi}^{2}$ profile corresponding to the redshift search for minimizing the distance between the L$_i^{nTh1}$--E$_{peak,i}^{nTh1,rest}$ data points for GRB 110205A (blue in panel (a)) and the BATSE L$_i^{nTh1}$--E$_{peak,i}^{nTh1,rest}$ relation (dashed black line in panel (a)). The spectroscopic (red line) and estimated redshifts are in agreement at the $\sim$1-$\sigma$ confidence level.}
\end{center}
\end{figure*}

C$_{nTh2}$'s fast temporal variability implies that it cannot arise from the outflowÕs interaction with the external medium~\citep{Granot:2011}. Its lack of temporal correlation with C$_{Th}$ and C$_{nTh1}$ suggests that it comes from a distinct emission region. We suggest a general scenario in which this can naturally occur. In this picture, the outflow magnetization near the central source, $\sigma_0$, varies with time, e.g., due to a fluctuating degree of mass entrainment from the sides of the GRB jet near its base. This would naturally lead to variations in $\sigma$ as a function of distance $R$ from the source, which would remain embedded in the flow to very large distances~\citep[even though $\sigma$ generally decreases with $R$ as the jet accelerates -- ][]{Asano:2009,Asano:2010}. This would naturally lead to different $\sigma$ values at the dissipation radius, $R_{dis}$~\citep[e.g.,][]{Drenkhahn:2002a,Drenkhahn:2002b,Tchekhovskoy:2008}. As a result, internal shocks would dominate in the low-$\sigma$ ($\sigma$$<$1) regions, leading to a  C$_{nTh1}$-like spectrum, whereas this dissipation channel is suppressed in high-$\sigma$ regions. A natural candidate that can efficiently dissipate magnetic energy in strongly magnetized flows is magnetic reconnection, which may also yield a C$_{nTh2}$-like spectrum. In this way the value of the magnetization parameter at the dissipation radius, $\sigma(R_{dis})$, determines the dominant dissipation mechanism---internal shocks or magnetic reconnection, which in turn determines the particle acceleration mechanism and the resulting emission spectrum. This effectively produces two distinct types of emission regions within the same GRB outflow.

The stochastic nature of magnetic reconnection that powers C$_{nTh2}$ naturally explains its lack of temporal correlation with C$_{nTh1}$ and C$_{Th}$, which more closely follow the central source activity. The origin of C$_{nTh2}$'s exact spectral shape is unclear, but it is viable given the large uncertainty on the exact physical conditions within a relativistic magnetic reconnection layer. The difference between the values of $\alpha_{nTh2}$ and $\alpha_{nTh1}$ is natural given their different dissipation and particle acceleration mechanisms.

\section{Toward a New Type of Standard Candles}

In summary, we have presented a unified spectro-temporal model for the broadband prompt emission, from the optical domain up $\gamma$-rays, which directly relates to the underlying dissipation and emission processes within their relativistic outflows. Our new L$_i^{nTh1}$--E$_{peak,i}^{nTh1,rest}$ luminosity-hardness relation intrinsic to C$_{nTh1}$~\citep{Guiriec:2013,Guiriec:2015a,Guiriec:2015b,Guiriec:2016} derived from GRBs with spectroscopic redshift measurements detected with CGRO/BATSE, also holds for GRB 110205A (Figure~\ref{fig04}a). Using our L$_i^{nTh1}$--E$_{peak,i}^{nTh1,rest}$ relation for BATSE data as a reference, we estimate the redshift of GRB 110205A to be z$=$2.5$\pm$0.2, well in agreement with its spectroscopic redshift of z$=$2.22, especially when one considers that BATSE, on one side, and Swift and Suzaku, on the other side, have never been inter-calibrated (Figure~\ref{fig04}b). Thus the confirmation of our new L$_i^{nTh1}$--E$_{peak,i}^{nTh1,rest}$ relation in GRB 110205A strongly supports the potential use of GRBs as cosmological standard candles; however, analysis of a larger GRB sample remains to be performed to confirm this result and estimate the accuracy of the redshift estimator.

\section{Acknowledgements}

We thank Takanori Sakamoto for the precious help. To complete this project, S.G. was supported by the NASA grants NNH11ZDA001N and NNH13ZDA001N, which were awarded to S.G. during cycles 5 and 7 of the NASA Fermi Guest Investigator Program. P.M. was supported by the NASA grant NNX13AH50G. J.G. and R.G. acknowledge support from the Israeli Science Foundation under Grant No. 719/14. R.G. is supported by an Outstanding Postdoctoral Researcher Fellowship at the Open University of Israel.

\end{document}